\documentclass{article}

\usepackage{arxiv}
\usepackage{float}
\usepackage[utf8]{inputenc} 
\usepackage[T1]{fontenc}    
\usepackage{hyperref}       
\usepackage{url}            
\usepackage{booktabs}       
\usepackage{amsfonts}       
\usepackage{nicefrac}       
\usepackage{microtype}      
\usepackage{lipsum}

\usepackage{graphicx}

\title{Designing compact training sets for data-driven molecular property prediction}

\author{
  Bowen Li,
 Srinivas Rangarajan\\
 \\
Department of Chemical and Biomolecular Engineering, Lehigh University,  Bethlehem
}

\begin{document}
\maketitle

\begin{abstract}

In this paper, we consider the problem of designing a training set using the most informative molecules from a specified library to build data-driven molecular property models. Specifically, we use (i) sparse generalized group additivity and (ii) kernel ridge regression as two representative classes of models, we propose a method combining rigorous model-based design of experiments and cheminformatics-based diversity-maximizing subset selection within the $\epsilon$--greedy framework to systematically minimize the amount of data needed to train these models. We demonstrate the effectiveness of the algorithm on subsets of various databases, including QM7, NIST, and a catalysis dataset. For sparse group additive models, a balance between exploration (diversity-maximizing selection) and exploitation (D-optimality selection) leads to learning with a fraction (sometimes as little as 15\%) of the data to achieve similar accuracy as five-fold cross validation on the entire set. On the other hand, kernel ridge regression prefers diversity-maximizing selections. 

\end{abstract}

\section{Introduction}

In several applications\cite{weber2011redox,ma2015machine,yu2012identification}, including high throughput screening for molecule discovery, modeling of large reaction systems, and design of catalysts, one encounters the challenge of estimating multiple properties of a large set of molecules (or materials, reaction intermediates, etc.). 
Experimental or ab initio estimation of such properties can be prohibitively expensive; a more computationally scalable strategy is, therefore, needed. 

Machine learning algorithms serve as an appealing alternative strategy for molecule property prediction. By training a surrogate model with reference data from quantum mechanical calculations or experiments, fast and accurate property prediction can be achieved. A number of molecular descriptors\cite{hansen2013assessment,hansen2015machine,rogers2010extended,rupp2015machine} and accurate machine learning models\cite{lusci2013deep,duvenaud2015convolutional,kearnes2016molecular} have been proposed for the property prediction of organic molecules. However, such techniques can be expensive because these machine learned models are often built using large datasets comprising 
between tens of thousands to millions of data points.

In this broader context, we consider the following question
: \textit{Given a library (or a set) of molecules, $\mathbf{S}$, for which we need to calculate properties, how does one identify the smallest subset of molecules $\mathbf{S_T}$, ($\mathbf{S_T} \subset \mathbf{S}$), to do experiments/computations on, such that a surrogate property model trained on $\mathbf{S_T}$ is accurate enough to be applied to the rest of the molecules in $\mathbf{S}$?} 

The concept of active learning or sequential experimental design aims to minimize the total number of training samples
for learning by adaptively selecting informative samples to be included in the training data.  This ultimately allows for building a relatively small but high-quality training set (relative to the design space) to save resources on data acquisition \cite{reker2017active,lang2016feasibility,reker2015active,liu2004active}. For molecular property prediction, active learning has been applied to achieve better accuracy than random selection with the same training set size \cite{tang2018prediction,gubaev2018machine}. 

In this work, we extend such a purely statistical approach by infusing domain information in an $\epsilon$--greedy fashion, reminiscent of reinforcement learning. In particular, the "domain knowledge" we use here is a diversity maximization algorithm to select molecules into $\mathbf{S_T}$. 
We consider the sequential experimental design as exploiting the model for the best short-term reward, and diversity maximization as the exploration of the molecule space to improve the model by introducing new fragments into the model. The ratio of exploitation vs exploration moves is set by a user-provided fraction $\epsilon$. As a proof-of-concept, we test this algorithm on three datasets of varying molecule sizes, types, and properties. Our property prediction models are based on a sparse version of a generalized group additivity formalism that offers a reasonably accurate and explainable data-driven model for molecular properties. For such models, we demonstrate the algorithm by showing that $\mathbf{S_T}$ can have less than $15-50\%$ of $\mathbf{S}$. Finally, we extend our algorithm to kernel-based property models where we see that diversity maximization results in better training sets. We begin with the discussion of methods.

\section{Methods}
\subsection{Sparse group additivity model building}

Group additivity (GA), proposed by Benson and Buss in 1958, is a simple linear property model where the property of interest is calculated as the sum of contributions coming from constituent atoms, bonds, or groups. \cite{benson1958additivity,benson1969additivity,eigenmann1973revised,cohen1993estimation}
\begin{equation}
f(x) = \sum_{i=1}^P \beta_ix_i
\label{eq:OLS}
\end{equation}
where $f(x)$ is enthalpy or entropy of formation, $x_i$ is for instance the number of groups and $\beta_i$ is the coefficient weight (or the contribution). 

Traditionally, groups were "handpicked" by the developer. As an alternative, one could use any of the more modern cheminformatics-based fingerprints. We use pathway fingerprints, which are modified from Daylight fingerprints (a Boolean array showing absence or presence of patterns in a molecule), by enumerating each atom and linear substructure. 
Specifically, paths of length one to seven (1—-7) atoms in the molecular graphs are identified. Subsequently, any molecule can be represented (not uniquely) by storing the count of each path (or essentially fragment) in a fingerprint vector.

Two additional correction terms for pathway fingerprints describing aromaticity and ring information are introduced since they are not covered in linear substructures. The detailed steps to build modified pathway fingerprints are contained in section S1 in supporting information.

The design matrix, $X$, is formed in which each row represents the fingerprints vector of a molecule in the dataset. GA can be applied for property prediction using this matrix. A common way for obtaining the feature coefficient $\beta_i$  would be ordinary least square (OLS) regression:  
\begin{equation}
\hat{\beta} = arg \, \min_{\beta} \, \left\|y-X\beta\right\|^2
\label{eq:2}
\end{equation}

The coefficient vector can be obtained from
\begin{equation}
\hat{\beta} = \left(X^TX\right)^{-1}X^Ty
\label{eq:coef}
\end{equation}

Pathway fingerprints can result in a large number of features. We obtain sparse models to prevent overfitting, we use Least Absolute Shrinkage and Selection Operator (LASSO), to pick relevant features by introducing a 1-norm regularization term with penalty $\lambda$ into the loss function of the OLS: 

\begin{equation}
\hat{\beta} = \min_{\beta \in \mathbb{R}^P} \left\{\frac{1}{N} \left\|y-X\beta\right\|^2_2 + \lambda\left\|\beta\right\|_1\right\}
\label{eq:LASSO}
\end{equation}

Thus a sparse solution, of dimension \textit{p}, is obtained by penalizing the coefficient weights and pushing insignificant ones to zero. The value of regularization term $\lambda$ relates to the sparsity of solution; a larger $\lambda$ results in less nonzero features. The optimal $\lambda$ value is chosen by minimizing a cross-validation error on the training set.

A similar approach utilizing exhaustively generated subgraphs of molecules and LASSO feature selection shows good performance on gaseous species and surface intermediates.\cite{gu2018thermochemistry}.

The GA model is built with exhaustively generated pathway fingerprints features and LASSO feature selection algorithm
. LASSO is performed with scikit-learn Elastic net python module by setting L2 norm to zero.

\subsection{Max-Min dissimilarity-based subset selection}

Structurally similar molecules are likely to exhibit similar properties\cite{johnson1990concepts}, thus $\mathbf{S_T}$ can contain more information related to property by maximizing the coverage of fragments in that set. This concept is widely applied to maximize diversity in drug discovery and design, which favors screening small yet diverse fragment-based libraries instead of the conventional strategy that aims to evaluate as many compounds as technologically possible\cite{hajduk2007decade,bures1998computational,maldonado2006molecular}.

In this work, we use the \textit{Max-Min} method for diversity maximization and S\o rensen Dice coefficient (DSC) for similarity quantification to select the most fragment-diverse  subset \cite{ashton2002identification}. S\o rensen Dice coefficient  converts the similarity into a real number by:

\begin{equation}
DSC = \frac{2\left|Y\cap Z\right|}{\left|Y\right|+\left|Z\right|}
\label{eq:DSC}
\end{equation}

where $Y$ and $Z$ are two vectors or sets, and the numerator of the coefficient is twice the number of common elements in both sets while the denominator is the sum of the number of elements in each set. The larger the DSC value between two sets the more similar they are.

The Max-Min method maximizes the fragment diversity by: (i) calculating the smallest dissimilarity between each compound in the remaining set and those already in the training set, and (ii) selecting the molecule with the largest dissimilarity to be added into $\mathbf{S_T}$.  The method essentially tries, in a greedy sense, to minimize the maximum distance between molecules outside $\mathbf{S_T}$ and within it; hence the term ``Max-Min". Note that, the smallest dissimilarity value is the largest  S\o rensen Dice coefficient and $vise \  versa$. Further, the algorithm can be employed \textit{a priori} (before calculating the desired properties) and can be stopped when $\mathbf{S_T}$ reaches a desired size.

In this work, when applying the Max-Min method for the diversity maximization of pathway fingerprints for a given set, the maximization is performed through the rdkit python module 'MaxMinPicker' and based on its
built-in pathway fingerprints module 'FingerprintMols', wherein the fingerprints are hashed and lumped to improve efficiency.

\subsection{D-optimal sequential experimental design}

When cases arise where samples (molecules) are abundant but their labels (properties) are not readily available or are expensive to acquire, active learning or experimental design methods \cite{cohn1996active} can be implemented to choose the most informative samples to label (i.e. to acquire property for a specific molecule). Active learning methods are supervised machine learning methods wherein the training data is updated adaptively to build a compact but high-quality model. Popular selection strategies include uncertainty sampling\cite{lewis1994heterogeneous,tong2001support}, optimal experimental design\cite{yu2006active}, query-by-committee\cite{seung1992query} and querying representative examples\cite{huang2010active}. Based on a particular strategy, information-rich samples are selected from the remaining samples and added into the current training set after obtaining their labels (properties); thus the model can be improved in an adaptive manner.

In OLS (Equation \ref{eq:2}), the $p \, \times p$ matrix $(X^TX)$ is called the information matrix of the parameters $\beta$; a larger the determinant of $(X^TX)$ translates to a more informative training set. The goal of optimal experimental design is to find the best combination of samples from the dataset to optimize different properties of information matrices based on experimental design criteria\cite{atkinson2007optimum}. The most commonly applied one is the D-optimality criterion\cite{smith1918standard}, which maximizes the determinant of the information matrix and is equivalent to minimizing the determinant of the covariance matrix $V = (X^TX)^{-1}$.

Geometrically, the related $(1-a)\% $ confidence region of the estimated parameters $\hat{\beta}$ of $p$ dimensions can be represented by the confidence ellipsoid obtained from the $F$ test which satisfies the following inequality\cite{draper1998applied}: 

\begin{equation}
\left(\beta-\hat{\beta}\right)^TX^TX\left(\beta-\hat{\beta}\right) \leq ps^2F_{p,v,a}
\label{eq:ellipsoid}
\end{equation}

where $\beta$ represents any point in the confidence region, $s^2$ is the estimate of variance $\sigma^2$, and $F$ is the $F$-statistic corresponding to $v$ degree of freedom, $p$ number of parameters and the $a$ level of significance. The axes of the confidence ellipsoid represent the confidence intervals of $\hat{\beta}$, thus shrinking the volume of the ellipsoid leads to reducing the parameter's errors. As the volume is proportional to the square root of the determinant of the covariance matrix $V$, the generalized variance of the estimated parameters $\hat{\beta}$ is minimized by performing the D-optimality criterion \cite{atkinson2007optimum}.

We note that for linear models, a D-optimal design of a desired number of training points can be identified \textit{a priori} (e.g. the DETMAX algorithm \cite{mitchell1974algorithm}). For nonlinear models and in the case of our sparse models wherein the matrix $X$ can change depending on which parameters (i.e. paths or fragments) are picked in each iteration, one can use a sequential design approach to greedily maximize the information content. 

We use the sequential D-optimal experimental design as one of the strategies to update the most informative remaining molecule into the model iteratively. At each iteration, we first calculate our sparse regression model to determine our matrix $X$; then, for each remaining molecule (in $\mathbf{S} - \mathbf{S_T}$) with a feature vector $x_i$, the covariance matrix is updated using the Sherman-Morrison formula\cite{sherman1950adjustment}

\begin{equation}
V' = V - \frac{Vx^TxV}{1+xVx^T}
\label{eq:SM}
\end{equation}

as a numerically inexpensive way to calculate the otherwise expensive inverse of the updated $X^TX$. Here, $V$ is the current covariance matrix and $V'$ is the updated covariance matrix. The molecule that lowers the determinant of the updated covariance matrix the greatest is considered to contain the most information, its property is then calculated and updated into the training set.

\subsection{Epsilon greedy method for exploration-exploitation trade-off}

Within the sequential experimental design, the training set is updated at each step with a molecule that is calculated to be the most informative. Such sequential updates are greedy actions which keep \textit{exploiting} the current model; therefore, can be biased by the model. One way to improve upon such purely exploitative updates is to allow the possibility of selecting samples to the training set that do not offer the best short-term expectations, but potentially offer long-term benefits of improved solutions (i.e. a lower error on the remaining set for a given training set size). Selecting such samples is called \textit{exploration}. Since at each step there is a choice of exploring the space or exploiting the current model, the problem is referred to as a ``conflict" between exploration and exploitation \cite{sutton2018reinforcement}.

Epsilon-greedy method is a widely used strategy to balance exploitation and exploration moves in reinforcement learning studies\cite{tokic2011value,wunder2010classes}, where the ratio of exploration is controlled by a hyperparameter $\epsilon$. At each step, a random number $\zeta$ is generated between 0 and 1, if $\zeta$ falls in a specified range determined by $\epsilon$, a random (or a more carefully chosen exploratory) action would be taken instead of the best exploitation action; otherwise, an exploitation action is chosen.

We add exploration actions into the sequential experimental design (which is the exploitation action) through the epsilon greedy strategy. For exploration, we use the Max-Min method to pick a molecule that maximizes fragment diversity in the training set. At each iteration, the action is taken according to the generated random number $\zeta$ to determine whether we should pick a molecule based on Max-Min method to increase the model feature diversity or based on  D-optimality to reduce the generalized variance of the estimated parameters. If $\zeta$ is smaller than $1-\epsilon+\frac{\epsilon}{n}$, where $n$ is the number of remaining molecules, the molecule will be updated using the  D-optimality strategy, otherwise, it is updated using the Max-Min method.

\subsection{Iterative optimal training set building algorithm}
\begin{figure*}[ht]
\centering
  \includegraphics[width=0.9\textwidth]{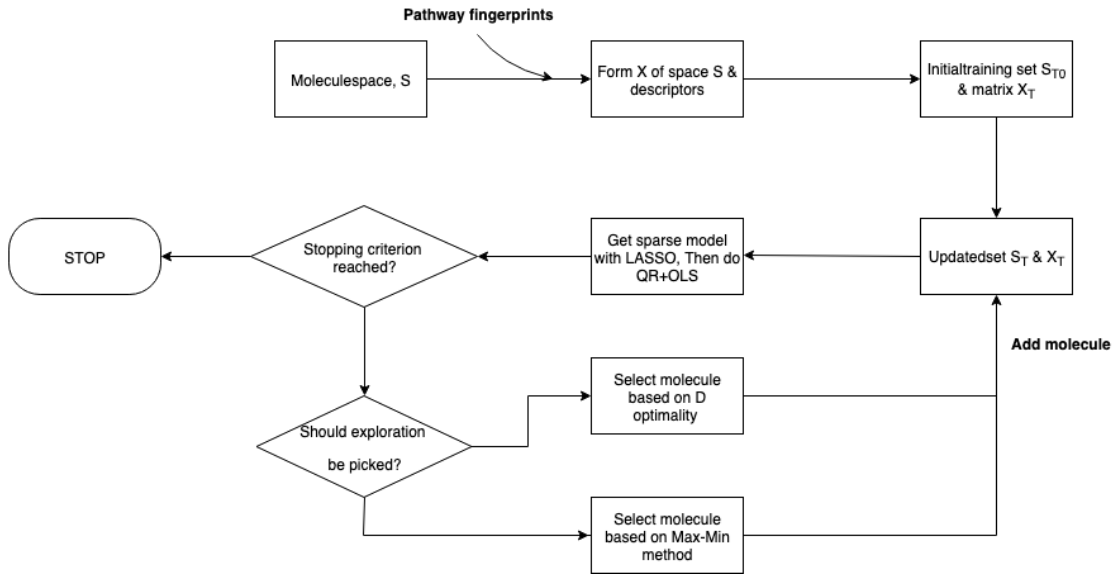}
\caption{Workflow of the proposed algorithm for building the optimal training set.
}
\label{flowchart}
\end{figure*}

The flowchart of our algorithm is shown in Figure \ref{flowchart}. Given a molecular space $\mathbf{S}$, the design matrix $\mathbf{X}$ is constructed with descriptors (all pathway fingerprints up to a specified length) to represent the molecules in the space. The algorithm starts with selecting molecules from $\mathbf{S}$ as the initial training set $\mathbf{S_{T0}}$  and forming initial training set matrix $\mathbf{X_T}$ built from $\mathbf{X}$ with each row representing selected molecule. The next step is to apply LASSO on $\mathbf{X_T}$ to select a subset of the columns; we further carry out a QR decomposition to remove linearly dependent columns. The model matrix can be transformed into the information matrix which is needed for implementing the D-optimal sampling. After LASSO and QR decomposition, an OLS step (with the reduced set of columns) is carried out to obtain a final model and its associated prediction (and standard error) for the remaining molecules in $\mathbf{S} - \mathbf{S_T}$. Subsequently, if the stopping criterion is not reached, the epsilon greedy strategy is applied to determine whether to perform an exploration step (using the Max-Min method) or an exploitation step (using D-optimality). Based on the selection strategy a molecule from the remaining set $\mathbf{S} - \mathbf{S_T}$ is selected and added into $\mathbf{S_T}$,  its property is obtained and its feature vector appended to $\mathbf{X_T}$. This iterative update of $\mathbf{S_T}$ and $\mathbf{X_T}$ occurs until the stopping criterion is reached. Note that LASSO at each iteration considers the whole set of P fingerprints from which it identifies p salient ones. A critical aspect of this algorithm is to decide when to stop; we discuss potential criteria in a subsequent section. The details of how to select the regularization value for LASSO in each step is discussed in section S2 in supporting information. 

\subsection{Training set building methods for kernel model}

A kernel model assumes that samples (molecules) with similar patterns are likely to have similar labels (properties). The prediction value of a sample is, therefore:

\begin{equation}
f(x) = \sum_{i=1}^n \alpha_ik\left\langle x_i,x \right\rangle
\label{eq:kernel}
\end{equation}

where $\alpha_i$ is the coefficient of each training sample representing its contribution and $k\left\langle x_i,x \right\rangle$ is the kernel function chosen to replace the inner product computation in feature space to save computation resources.

We test both diversity maximization and variance sampling selection strategies for the kernel model. For maximizing diversity, we use the Max-Min method to minimize the maximum kernel distance between molecules within the training set and those outside. We use variance sampling, similar to the uncertainty sampling in classification, that directly aims to select the molecule with the largest prediction variance. The prediction variance is calculated by ${k_i^{T}(K^TK)^{-1}k_i}$ based on kernel matrix $K$ of the current training set and kernel distance vector $k_i$ with each element representing the kernel distance between the remaining molecule $i$ and each training molecule. This is similar to the sequential D-optimal design.

The model is based on the Laplacian kernel and the hyperparameter  $\sigma$ (kernel width) is set for the entire exercise as max$\left(\left\{D_{ij}\right)\right\}/ln\left(2\right)$ where $D_{ij}$ is the kernel distance of each pair of molecules in the initial training set. The regularization term $\lambda$ is set to $0$ following the arguments of   Ramakrishnan.\cite{ramakrishnan2015many} Bag of bonds (BOB) \cite{hansen2015machine} molecular representation is used to quantify the difference of two molecules with their 3D coordinates as inputs.

\section{Datasets}

Our method is benchmarked on three different sets of data: (1)$\sim$7000 molecules consisting of C,O,N,S,H from the QM7 database\cite{blum2009970,rupp2012fast}, (2) 920 hydrocarbon species from NIST chemistry webbook\cite{buerger2017big,Linstrom2005nist}
,  and (3) 591 surface intermediates on a transition metal facet \cite{gu2016group}. The QM7 dataset contains atomization energies (computed using density functional theory, or DFT) of molecules with up to seven (7) non-H atoms. The NIST chemistry webbook contains hydrocarbon molecules and their experimental heats of formation values; the size of the molecules are more diverse than QM7 dataset as the largest molecule contains 42 heavy atoms. The ``surface intermediates" dataset contains heat of formation values calculated using DFT for different surface adsorbates involved in the conversion of lignin monomers. We apply pathway fingerprints as the descriptor on the first two datasets and use the adsorbate graph mining algorithm\cite{gu2018thermochemistry} to subtract groups from adsorbates for the third case.

Given that pathway fingerprints are linear substructures, molecules with multiple ring fragments (e.g. fused rings) are difficult to accurately capture. Thus the performance of our algorithm is benchmarked on the subsets of QM7 dataset and NIST chemistry webbook by excluding about 600 and 331 molecules respectively with more than 1 ring fragment. 

\section{Results and Discussion}

The performance of a selection strategy is measured by the root mean squared error (RMSE) of the prediction on the remaining samples with the model trained on selected samples ($\mathbf{S_T}$). For each set of data, the average RMSE of the five-fold cross validation (CV) is shown as a baseline to represent the performance of the model with sufficiently large training molecules. We note that the RMSE of the remaining set is not theoretically bounded by this baseline; nevertheless, for a large $\mathbf{S}$ and a relatively small $\mathbf{S_T}$, the baseline is a worthy target for the models. The relationship between the values of the regularization and the corresponding RMSE and MAE is shown in section S3 in the supporting information (SI).

For any step that involves random number generation, we report the averages of ten different runs. 
This then leads to averaging over (1) hundred runs for random sampling (ten different sets for the initial set and ten simulations of updates for each initial set), (2) ten runs for the D-optimal strategy each starting with a different random initial training set, (3) ten runs for largest error elimination curve, and (4) ten runs for each $\epsilon$ value for the epsilon greedy method. 

\subsection{Comparison of selection strategies for the GA model on QM7}

\begin{figure*}[hbt!]
\centering
\includegraphics[width=0.6\textwidth]{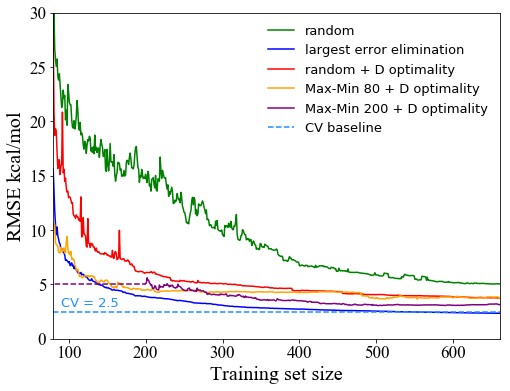}
\caption{Comparison of the learning rates of different molecule selection strategies on the QM7 dataset for GA-based prediction of heats of atomization. The y-axis measures the root mean square error of the prediction on the remaining set during the updating process with (1) random sampling, (2) largest error elimination method, (3) D-optimality strategy with initial training set selected randomly or by Max-Min method. The baseline represents the five-fold cross validation result of the dataset. The reported performance of random sampling, largest error elimination method and the D-optimality strategy with initial training set selected randomly are averaged over multiple runs}
\label{evolve}
\end{figure*} 

For the QM7 dataset, we first set the bounds for the selection strategies - (i) the worst case is  when the remaining set ($\mathbf{S} - \mathbf{S_T}$) is sampled randomly to pick the best molecule to $\mathbf{S_T}$ and (ii) the best case is denoted as the "largest error elimination method" wherein all labels (properties) are known and in each iteration we select the molecule with the largest prediction error (the difference between the true value and the predicted value using the model trained on $\mathbf{S_T}$) from the optimal model.

The performance of the two strategies is shown in Figure \ref{evolve}. The two selection strategies start with the same randomly chosen molecules (of size eighty) as the initial training set (generated with the same random seed). Then we evaluate their performance by inspecting the predicted RMSE of the remaining set as a function of the size of $\mathbf{S_T}$. Compared to random sampling, the largest error elimination strategy, as expected, rapidly converges to the CV baseline. In practice, however, all properties are not available and the largest error elimination strategy is purely a theoretical exercise; nevertheless, the gap in the learning rate (the rate of decrease of RMSE with increasing size of the training set) between the two strategies suggests that, in principle, a model trained on a carefully selected small subset can perform as good as the one trained on the entire set.

Next, we investigate the performance of the D-optimality strategy, which is shown in red in Figure \ref{evolve}. Clearly, the strategy learns faster than random sampling and is able to reach the CV baseline within 1.5 kcal/mol with just 600 molecules in the training set. However, even though the D-optimal learning rate is high early on, its performance starts to slow down at about 300 molecules. This indicates that the initial training set does not have sufficient representation of salient features of the QM7 dataset; the D-optimal selection strategy then uses a statistical criterion to add new molecules into the training set. While this maximizes the determinant of the information matrix it does not ensure that relevant missing features are included in the training set.    

To investigate whether missing relevant features can be identified by picking appropriate molecules, we next apply the Max-Min method to build the initial training set for the D-optimal selection strategy (``Max-Min 80 + D optimality" curve in Figure \ref{evolve}). We can observe that early on this curve outperforms the original D-optimal curve and learns almost as fast as the largest error elimination curve. If we further increase the number of initial training set molecules from 80 to 200 (shown in the "Max-Min 200 + D-optimality" curve in Figure \ref{evolve}), the curve approaches the baseline and is within 0.5 kcal/mol for $|\mathbf{S_T}|>500$. The result indicates that the performance of the D-optimal strategy improves upon starting with a diverse training set.

\subsection{Epsilon Greedy Method}

\begin{figure*}[hbt!]
\centering
\includegraphics[width=0.6\textwidth]{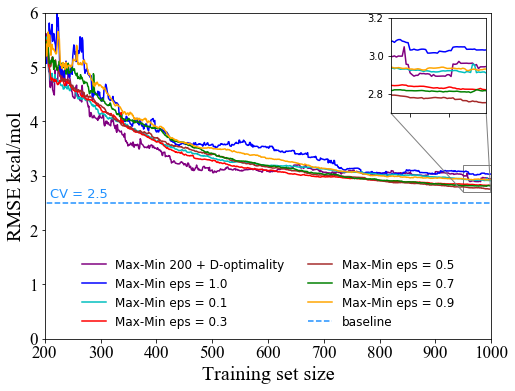}
\caption{Comparison of the learning rate of different $\epsilon$ (eps) values on the QM7 dataset for GA-based prediction of heats of atomization. The value of $\epsilon$ determines the ratio of sample selection with the D-optimal criterion (exploitation) or Max-Min diversity maximization (exploration). The reported performance of strategies combining both exploration and exploitation ($\epsilon$ between 0.1 and 0.9) are averaged over ten runs.
}
\label{eps_method}
\end{figure*} 

Introducing diversity maximization to identify the initial training set improves the learning rate of the D-optimal strategy. Therefore, it is clear that the Max-Min algorithm is able to explore the molecule space better than random selection. 
In Figure \ref{eps_method} we show the performance of the epsilon-greedy method to combine the D-optimal strategy and Max-min methods for molecule selection. The figure shows different values of the hyperparameter $\epsilon$ in order to identify the optimal one. $\epsilon$ values range from 0 (purely exploitative using D-optimal) to 1 (purely explorative with Max-Min). Figure \ref{eps_method} shows that the D-optimality curve ($\epsilon=0$) performs the best in the early stages but the advantage vanishes after about 600 molecules, while the pure Max-Min method ($\epsilon=1$) decreases the slowest. The learning curve for $\epsilon =0.5$ lies between those with the pure methods ($\epsilon=0/1$)  in the early steps and yields the best ``long-term" performance (by about 0.2 - 0.3 kcal/mol when the training set size reaches 1000). The results show that exploitation provides a good short-term benefit while the exploration improves the model in the long run. A balance between exploration and exploitation strategies can lead to a steady and on overall better performance. The detailed values for the comparison of the selection strategies are reported in Table \ref{table_accuracy}.

\begin{table}[hbt!]
\centering
\caption{Comparison of the RMSE, the largest prediction standard error and the average prediction standard error of the remaining set with different training set sizes and different selection strategies for GA-based prediction for heats of atomization (kcal/mol) of the QM7 dataset. The number in the parenthesis of each strategy shows the size of initial training set to start with. The performance of strategies except Max-Min + D-optimality (200) are averaged from multiple runs. The largest prediction standard error and the average prediction standard error for random sampling and largest error elimination method are not reported.}
\label{table_accuracy}
\resizebox{.9\textwidth}{!}{%
\begin{tabular}{ccccccccc}
\hline
\begin{tabular}[c]{@{}l@{}}Training \\ set size\end{tabular} & \multicolumn{2}{c}{Random (80)}                                  & \multicolumn{2}{c}{\begin{tabular}[c]{@{}c@{}}Largest error \\ elimination (80)\end{tabular}} & \multicolumn{4}{c}{\begin{tabular}[c]{@{}c@{}}Random + \\ D-optimality (80)\end{tabular}}                                                                                                   \\ \hline
                                                             & RMSE  & \begin{tabular}[c]{@{}l@{}}Training\\\hfil error\end{tabular} & RMSE                & \begin{tabular}[c]{@{}l@{}}Training\\\hfil error\end{tabular}                & RMSE & \begin{tabular}[c]{@{}l@{}}Training\\\hfil error\end{tabular} & \begin{tabular}[c]{@{}l@{}}\hfil Largest \\ standard error\end{tabular} & \begin{tabular}[c]{@{}l@{}}\hfil Average \\ standard error\end{tabular} \\ \hline
200                                                          & 15.13 & 3.07                                                     & 3.88                & 6.34                                                                    & 6.08 & 2.66                                                     & 4.81                                                        & 2.21                                                       \\
600                                                          & 5.14  & 2.50                                                     & 2.41                & 6.24                                                                    & 3.87 & 3.07                                                     & 2.30                                                       & 1.26                                                        \\
800                                                          &  --     &    --                                                      &    --                 &     --                                                                    & 3.52 & 3.19                                                     & 1.49                                                        & 1.06                                                       \\
1000                                                         &  --     &   --                                                       &   --                  &   --                                                                      & 3.36 & 3.32                                                     & 1.56                                                        & 0.94                                                        \\ \hline
\end{tabular}%
}
\end{table}
\begin{table}[hbt!]
\centering
\resizebox{.9\textwidth}{!}{%
\begin{tabular}{ccccccccc}
\hline
\begin{tabular}[c]{@{}l@{}}Training \\ set size\end{tabular} & \multicolumn{4}{c}{\begin{tabular}[c]{@{}c@{}}Max-Min + \\ D-optimality (200)\end{tabular}}                                                                                                 & \multicolumn{4}{c}{\begin{tabular}[c]{@{}c@{}}Epsilon greedy :\\  $\epsilon$ = 0.5 (200)\end{tabular}}                                                                                      \\ \hline
                                                             & RMSE & \begin{tabular}[c]{@{}l@{}}Training\\\hfil error\end{tabular} & \begin{tabular}[c]{@{}l@{}}   \hfil Largest \\ standard error\end{tabular} & \begin{tabular}[c]{@{}l@{}}\hfil Average \\ standard error\end{tabular} & RMSE & \begin{tabular}[c]{@{}l@{}}Training\\ \hfil error\end{tabular} & \begin{tabular}[c]{@{}l@{}}\hfil Largest \\ standard error\end{tabular} & \begin{tabular}[c]{@{}l@{}}\hfil Average \\ standard error\end{tabular} \\ \hline
200                                                          & 5.07 & 0.97                                                     & 15.90                                                       & 3.32                                                        & 5.07 & 0.97                                                     & 15.90                                                      & 3.32                                                        \\
600                                                          & 3.08 & 2.07                                                     & 1.97                                                        & 1.30                                                       & 3.16 & 2.24                                                     & 2.32                                                       & 1.31                                                        \\
800                                                          & 3.05 & 2.36                                                     & 1.56                                                        & 1.07                                                        & 2.90 & 2.41                                                     & 1.96                                                      & 1.13                                                        \\
1000                                                         & 2.94 & 2.58                                                     & 1.41                                                       & 0.97                                                        & 2.75 & 2.54                                                     & 1.66                                                       & 1.00                                                       \\ \hline
\end{tabular}%
}
\end{table}

\subsection{Applying the algorithm to the NIST set}

\begin{figure*}[hbt!]
\centering
\includegraphics[width=0.6\textwidth]{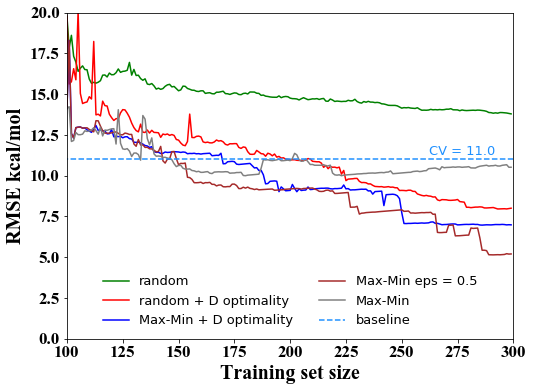}
\caption{Comparison of the learning rates of different molecule selection strategies on the NIST chemistry webbook for GA-based prediction of heats of formation. The selection strategy involves (1) random sampling, (2) D-optimality strategy with initial training set selected randomly or by Max-Min method, (3) epsilon greedy method with $\epsilon$ set to 0.5, (4) pure Max-Min method. The baseline represents the five-fold cross validation result of the dataset. The performance of random sampling, the D-optimality with initial training set selected randomly and epsilon greedy method is averaged over multiple runs.
}
\label{NIST_active_learning}
\end{figure*} 

We apply these different sampling strategies on $\sim$600 hydrocarbon molecules from the  NIST database to investigate whether the algorithm can perform well on a smaller experimental dataset comprising of diverse molecule size. Figure \ref{NIST_active_learning} shows the relative performance of different strategies. We note here that all methods eventually have a lower RMSE value than the "baseline" cross validation error, which is expected for a small set of molecules. Overall, the results are similar to that of the QM7 dataset: $\epsilon$-greedy with $\epsilon=0.5$ is better than D-optimality with Max-min initial selection, which in turn is better than purely random sampling. However, subtle differences arise; for instance, the pure Max-Min sampling performs well in the early stages but slows down after $\sim$200 molecules in the training set. Nevertheless, a combination of D-optimality and Max-Min method gives the best performance; this strategy approaches the CV base with $\sim$140 molecules in the training set (compared to $\sim$470 training molecules in each CV model) and reaches 5.2 kcal/mol with $\sim$300 training molecules(compared to 11.0 kcal/mol as the CV error). 

\subsection{Surface intermediates dataset}

\begin{figure*}[hbt!]
\centering
\includegraphics[width=0.6\textwidth]{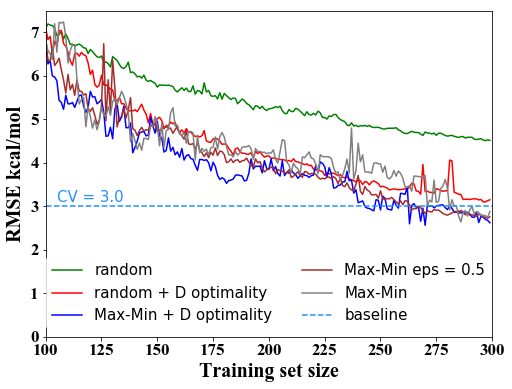}
\caption{Comparison of the learning rates of different molecule selection strategies on the surface intermediates dataset for GA-based prediction of heats of formation. The selection strategy involves (1) random sampling, (2) D-optimality strategy with initial training set selected randomly or by Max-Min method, (3) epsilon greedy method with $\epsilon$ set to 0.5, (4) pure Max-Min method. The baseline represents the five-fold cross validation result of the dataset. The reported performance of random sampling, the D-optimality strategy with an initial training set selected randomly and epsilon greedy method is averaged over multiple runs. 
}
\label{VGA_active_learning}
\end{figure*} 

The selection strategies are further tested using a dataset of surface intermediates to investigate their performance based on a different descriptor (groups mined from surface intermediates) and a different type of molecules (surface intermediates). Figure \ref{VGA_active_learning} shows that despite the changes in the descriptor and the type of molecules, the "active learning" algorithm remains effective; the selection strategies involving both Max-Min method and D-optimality ( "Max-Min + D optimality" curve and "Max-Min eps = 0.5" curve) yields overall the best performance and reaches the CV baseline (involving $\sim$ 450 training molecules) with  40-50\% less training molecules.

\subsection{Comparison of selecting strategies for the kernel model}

\begin{figure*}[ht]
\centering
\includegraphics[width=0.6\textwidth]{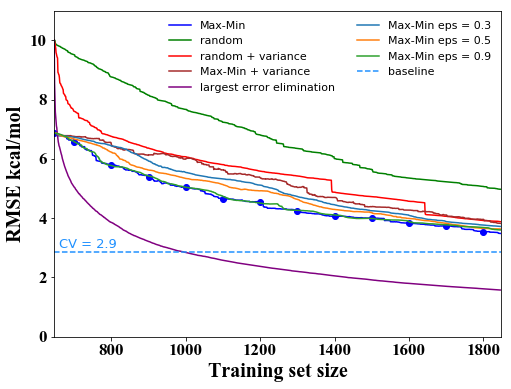}
\caption{Comparison of the learning rates of different molecule selection strategies on the QM7 dataset for kernel-based prediction of heats of atomization with bag of bonds descriptor. The selection strategies involve (1) random sampling, (2) largest error elimination method, (3) variance sampling with initial training molecules selected randomly or by Max-Min method, (4) the pure Max-Min method, (5) Epsilon greedy method with different $\epsilon$ values. The baseline represents the five-fold cross validation result of the dataset. The reported performance of random sampling, largest error elimination method, variance sampling with initial training molecules selected randomly and epsilon greedy method is averaged over multiple runs.
}
\label{bob}
\end{figure*}

Finally, we investigate the best selection strategy to build kernel models.  Figure \ref{bob} shows the result for the kernel model trained on the QM7 dataset using the ``bag of bonds" descriptor. Note that we apply a variance sampling for the ``exploitation" step, which is akin to the D-optimal strategy; Max-Min method uses kernel distance to quantify intermolecular similarity. Random sampling and largest error elimination continue to represent the worst and best cases respectively. Variance sampling started with a randomly selected initial training set outperforms random sampling, but the learning rate is not steady and the RMSE drops almost instantly when the training set size reaches $\sim$ 1400 and $\sim$ 1650. The reason is that out of the 10 executions (over which we average), while many perform well, a few suffer from failing to find one type of molecule that potentially represents a group of molecules in the space, which leads to the RMSE being essentially flat from the outliers. (Figure for the 10 executions is included in section S4 in supporting information.) On the other hand, the Max-Min method gives the best performance. This is expected since the optimal sampling strategy for kernel models is ensuring sufficient coverage of the molecule space. The learning curves of strategies that combine Max-Min method and variance sampling via epsilon greedy method lie between the pure methods ( "Max-Min + variance" curve and pure "Max-Min" curve). The higher the $\epsilon$ value is, $viz$ the more times the Max-Min method is applied, the performance is closer to the Max-Min method. The result indicates that for similarity-based property prediction models (i.e. using kernels), Max-Min method that maximizes kernel distance diversity is preferable.

\subsection{Stopping criterion}

\begin{figure*}[hbt!]
\centering
\includegraphics[width=1.0\textwidth]{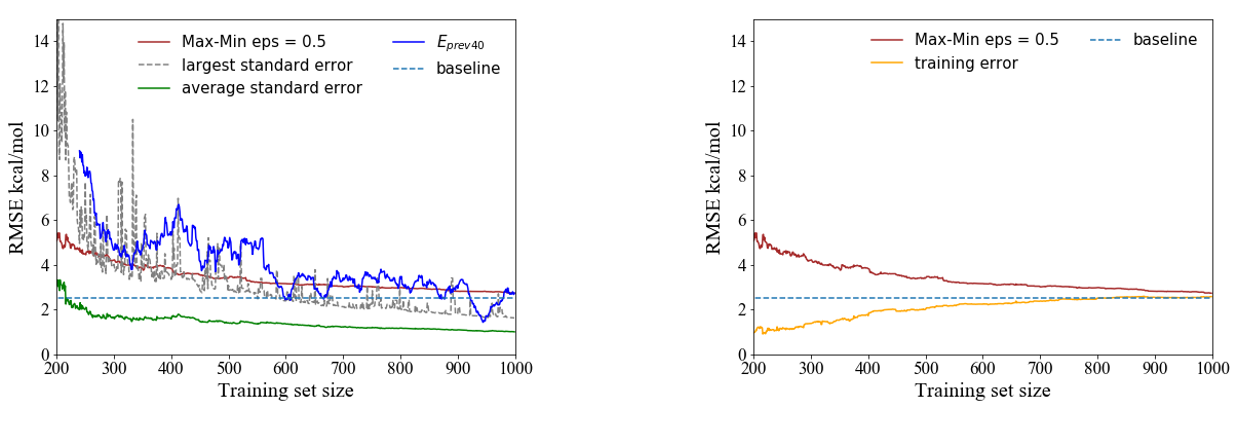}
\caption{Stopping criterion for the epsilon greedy method ($\epsilon = 0.5)$ applying on the GA model to predict heat of atomization for QM7 dataset. (left) The RMSE, the largest prediction standard error, and the average prediction standard error of the remaining set are shown. At the end of each iteration, the prediction error of the selected molecule can be calculated; $E_{prev40}$ is the moving average of this error over the previous forty (40) iterations. (right) Comparison of the RMSE of the remaining set and training set error.
}
\label{stop}
\end{figure*} 

It is essential to determine when to stop the algorithm. Although the prediction error on the remaining set is a useful metric to understand the performance of our method, it is not practical as labels are not available. Instead, we compute the prediction standard error as a surrogate to track the quality of the model and determine when the learning rate slows down. 

The prediction standard error of a molecule outside the training set is calculated as:

\begin{equation}
s = \sigma \sqrt{x_i^T\left(X^TX\right)^{-1}x_i}
\label{eq:st_d}
\end{equation}

where $x_i$ is the molecule feature vector and $X$ is the design matrix. 

In Figure \ref{stop} we first investigate whether the prediction standard error follows a similar trend with the model prediction accuracy, thereby allowing for its use as a surrogate. We compare the largest prediction standard error and the average prediction standard error of the remaining dataset with the RMSE at each iteration of the epsilon greedy strategy (eps = 0.5). We can see that as the two prediction standard error curves (``largest standard error" and ``average standard error") decrease rapidly in the early iterations, the model learns fast ( shown as ``Max-Min $\epsilon$ = 0.5" curve). As the prediction standard error curves are converging in the later stage, the improvement in accuracy is negligible. The similar behavior of the RMSE and the average standard error indicates that we can rely on the trend of the latter to qualitatively infer the learning rate. In principle, i) since the value of the largest prediction standard error in the remaining dataset reflects how much information we can gain by performing careful selection of the informative molecules, and ii) the average prediction standard error represents how confidently can the model predict the remaining molecules, one feasible stopping point can be when both of them are smaller than a user defined threshold,  

Further, we investigate if the quality of the model can be estimated by tracking the training error directly ( ``training error" curve) or the moving average of the difference between predicted and true values of the selected molecules in the previous forty iterations (denoted as $E_{prev40}$). Interestingly, the training error increases as more molecules are added indicating that the model overfits the data early on. But as more diverse data is added to the training set, the model is getting progressively less biased. Eventually with $\sim$900 training molecules selected, it reaches to the accuracy of the prediction on the remaining molecules. The result indicates that we may even use the training set error to estimate the prediction accuracy in the later stage of iteration. The ``$E_{prev40}$" also eventually drops to the RMSE value of the remaining set and the baseline error; however, it fluctuates considerably, although it could still be used as a qualitative metric.

\section{Conclusion}

In this work, we present an optimal training set building algorithm that identifies the smallest amount of informative molecules to be calculated to build a surrogate property training model $\mathbf{S_T}$ from a given molecule library $\mathbf{S}$ to predict on the remaining molecules $\mathbf{S} - \mathbf{S_T}$. The algorithm iteratively updates the sparse model generated with LASSO feature selection from the design matrix $\mathbf{X_T}$ with informative molecules selected based on either D-optimality experimental design as exploitation to improve the current model or Max-Min diversity maximization as exploration to introduce diversity into $\mathbf{X_T}$. The balance of the exploration and exploitation is controlled by the $\epsilon$ greedy method which is widely used in reinforcement learning studies. We validate the algorithm by comparing different selection strategies on three different datasets that vary in the size, type and properties of molecules: (1) gaseous molecules from QM7 dataset, (2) hydrocarbons from NIST chemistry webbook and (3) surface intermediates on transition metal catalysis. In all cases the algorithm that balances D-optimality and Max-Min method via the epsilon greedy method gives the best result 
and reduces the training molecules needed to $\sim$15\% of QM7 dataset,$\sim$25\% of NIST chemistry webbook and $\sim$45\% of surface intermediates dataset to achieve similar accuracy as their five-fold cross validation results. We extend our algorithm on a kernel model and show that building the training set with diversity maximization based on kernel distance is preferred. The GA and the kernel models in this work serve as two simple representation models for validation; in principle we can apply the algorithm to 
any property prediction model and fingerprints that are either user-defined or machine-learned.

\bibliographystyle{unsrt}  


\begin{thebibliography}{10}

\bibitem{weber2011redox}
Adam~Z Weber, Matthew~M Mench, Jeremy~P Meyers, Philip~N Ross, Jeffrey~T
  Gostick, and Qinghua Liu.
\newblock Redox flow batteries: a review.
\newblock {\em Journal of Applied Electrochemistry}, 41(10):1137, 2011.

\bibitem{ma2015machine}
Xianfeng Ma, Zheng Li, Luke~EK Achenie, and Hongliang Xin.
\newblock Machine-learning-augmented chemisorption model for co2
  electroreduction catalyst screening.
\newblock {\em The journal of physical chemistry letters}, 6(18):3528--3533,
  2015.

\bibitem{yu2012identification}
Liping Yu and Alex Zunger.
\newblock Identification of potential photovoltaic absorbers based on
  first-principles spectroscopic screening of materials.
\newblock {\em Physical review letters}, 108(6):068701, 2012.

\bibitem{hansen2013assessment}
Katja Hansen, Gr{\'e}goire Montavon, Franziska Biegler, Siamac Fazli, Matthias
  Rupp, Matthias Scheffler, O~Anatole Von~Lilienfeld, Alexandre Tkatchenko, and
  Klaus-Robert Mu?ller.
\newblock Assessment and validation of machine learning methods for predicting
  molecular atomization energies.
\newblock {\em Journal of Chemical Theory and Computation}, 9(8):3404--3419,
  2013.

\bibitem{hansen2015machine}
Katja Hansen, Franziska Biegler, Raghunathan Ramakrishnan, Wiktor Pronobis,
  O~Anatole Von~Lilienfeld, Klaus-Robert Mu?ller, and Alexandre Tkatchenko.
\newblock Machine learning predictions of molecular properties: Accurate
  many-body potentials and nonlocality in chemical space.
\newblock {\em The journal of physical chemistry letters}, 6(12):2326--2331,
  2015.

\bibitem{rogers2010extended}
David Rogers and Mathew Hahn.
\newblock Extended-connectivity fingerprints.
\newblock {\em Journal of chemical information and modeling}, 50(5):742--754,
  2010.

\bibitem{rupp2015machine}
Matthias Rupp.
\newblock Machine learning for quantum mechanics in a nutshell.
\newblock {\em International Journal of Quantum Chemistry}, 115(16):1058--1073,
  2015.

\bibitem{lusci2013deep}
Alessandro Lusci, Gianluca Pollastri, and Pierre Baldi.
\newblock Deep architectures and deep learning in chemoinformatics: the
  prediction of aqueous solubility for drug-like molecules.
\newblock {\em Journal of chemical information and modeling}, 53(7):1563--1575,
  2013.

\bibitem{duvenaud2015convolutional}
David~K Duvenaud, Dougal Maclaurin, Jorge Iparraguirre, Rafael Bombarell,
  Timothy Hirzel, Al{\'a}n Aspuru-Guzik, and Ryan~P Adams.
\newblock Convolutional networks on graphs for learning molecular fingerprints.
\newblock In {\em Advances in neural information processing systems}, pages
  2224--2232, 2015.

\bibitem{kearnes2016molecular}
Steven Kearnes, Kevin McCloskey, Marc Berndl, Vijay Pande, and Patrick Riley.
\newblock Molecular graph convolutions: moving beyond fingerprints.
\newblock {\em Journal of computer-aided molecular design}, 30(8):595--608,
  2016.

\bibitem{reker2017active}
Daniel Reker, Petra Schneider, Gisbert Schneider, and JB~Brown.
\newblock Active learning for computational chemogenomics.
\newblock {\em Future medicinal chemistry}, 9(4):381--402, 2017.

\bibitem{lang2016feasibility}
Tobias Lang, Florian Flachsenberg, Ulrike von Luxburg, and Matthias Rarey.
\newblock Feasibility of active machine learning for multiclass compound
  classification.
\newblock {\em Journal of chemical information and modeling}, 56(1):12--20,
  2016.

\bibitem{reker2015active}
Daniel Reker and Gisbert Schneider.
\newblock Active-learning strategies in computer-assisted drug discovery.
\newblock {\em Drug discovery today}, 20(4):458--465, 2015.

\bibitem{liu2004active}
Ying Liu.
\newblock Active learning with support vector machine applied to gene
  expression data for cancer classification.
\newblock {\em Journal of chemical information and computer sciences},
  44(6):1936--1941, 2004.

\bibitem{tang2018prediction}
Yu-Hang Tang and Wibe~A de~Jong.
\newblock Prediction of atomization energy using graph kernel and active
  learning.
\newblock {\em arXiv preprint arXiv:1810.07310}, 2018.

\bibitem{gubaev2018machine}
Konstantin Gubaev, Evgeny~V Podryabinkin, and Alexander~V Shapeev.
\newblock Machine learning of molecular properties: Locality and active
  learning.
\newblock {\em The Journal of Chemical Physics}, 148(24):241727, 2018.

\bibitem{benson1958additivity}
Sidney~W Benson and Jerry~H Buss.
\newblock Additivity rules for the estimation of molecular properties.
  thermodynamic properties.
\newblock {\em The Journal of Chemical Physics}, 29(3):546--572, 1958.

\bibitem{benson1969additivity}
Sidney~W Benson, FR~Cruickshank, DM~Golden, Gilbert~R Haugen, HE~O'neal,
  AS~Rodgers, Robert Shaw, and R~Walsh.
\newblock Additivity rules for the estimation of thermochemical properties.
\newblock {\em Chemical Reviews}, 69(3):279--324, 1969.

\bibitem{eigenmann1973revised}
HK~Eigenmann, DM~Golden, and SW~Benson.
\newblock Revised group additivity parameters for the enthalpies of formation
  of oxygen-containing organic compounds.
\newblock {\em The Journal of Physical Chemistry}, 77(13):1687--1691, 1973.

\bibitem{cohen1993estimation}
N~Cohen and SW~Benson.
\newblock Estimation of heats of formation of organic compounds by additivity
  methods.
\newblock {\em Chemical Reviews}, 93(7):2419--2438, 1993.

\bibitem{gu2018thermochemistry}
Geun~Ho Gu, Petr Plechac, and Dionisios~G Vlachos.
\newblock Thermochemistry of gas-phase and surface species via lasso-assisted
  subgraph selection.
\newblock {\em Reaction Chemistry \& Engineering}, 3(4):454--466, 2018.

\bibitem{johnson1990concepts}
Mark~A Johnson and Gerald~M Maggiora.
\newblock {\em Concepts and applications of molecular similarity}.
\newblock Wiley, 1990.

\bibitem{hajduk2007decade}
Philip~J Hajduk and Jonathan Greer.
\newblock A decade of fragment-based drug design: strategic advances and
  lessons learned.
\newblock {\em Nature reviews Drug discovery}, 6(3):211, 2007.

\bibitem{bures1998computational}
Mark~G Bures and Yvonne~C Martin.
\newblock Computational methods in molecular diversity and combinatorial
  chemistry.
\newblock {\em Current opinion in chemical biology}, 2(3):376--380, 1998.

\bibitem{maldonado2006molecular}
Ana~G Maldonado, JP~Doucet, Michel Petitjean, and Bo-Tao Fan.
\newblock Molecular similarity and diversity in chemoinformatics: from theory
  to applications.
\newblock {\em Molecular diversity}, 10(1):39--79, 2006.

\bibitem{ashton2002identification}
Mark Ashton, John Barnard, Florence Casset, Michael Charlton, Geoffrey Downs,
  Dominique Gorse, John Holliday, Roger Lahana, and Peter Willett.
\newblock Identification of diverse database subsets using property-based and
  fragment-based molecular descriptions.
\newblock {\em Quantitative Structure-Activity Relationships}, 21(6):598--604,
  2002.

\bibitem{cohn1996active}
David~A Cohn, Zoubin Ghahramani, and Michael~I Jordan.
\newblock Active learning with statistical models.
\newblock {\em Journal of artificial intelligence research}, 4:129--145, 1996.

\bibitem{lewis1994heterogeneous}
David~D Lewis and Jason Catlett.
\newblock Heterogeneous uncertainty sampling for supervised learning.
\newblock In {\em Machine Learning Proceedings 1994}, pages 148--156. Elsevier,
  1994.

\bibitem{tong2001support}
Simon Tong and Daphne Koller.
\newblock Support vector machine active learning with applications to text
  classification.
\newblock {\em Journal of machine learning research}, 2(Nov):45--66, 2001.

\bibitem{yu2006active}
Kai Yu, Jinbo Bi, and Volker Tresp.
\newblock Active learning via transductive experimental design.
\newblock In {\em Proceedings of the 23rd international conference on Machine
  learning}, pages 1081--1088. ACM, 2006.

\bibitem{seung1992query}
H~Sebastian Seung, Manfred Opper, and Haim Sompolinsky.
\newblock Query by committee.
\newblock In {\em Proceedings of the fifth annual workshop on Computational
  learning theory}, pages 287--294. ACM, 1992.

\bibitem{huang2010active}
Sheng-Jun Huang, Rong Jin, and Zhi-Hua Zhou.
\newblock Active learning by querying informative and representative examples.
\newblock In {\em Advances in neural information processing systems}, pages
  892--900, 2010.

\bibitem{atkinson2007optimum}
Anthony Atkinson, Alexander Donev, and Randall Tobias.
\newblock {\em Optimum experimental designs, with SAS}, volume~34.
\newblock Oxford University Press, 2007.

\bibitem{smith1918standard}
Kirstine Smith.
\newblock On the standard deviations of adjusted and interpolated values of an
  observed polynomial function and its constants and the guidance they give
  towards a proper choice of the distribution of observations.
\newblock {\em Biometrika}, 12(1/2):1--85, 1918.

\bibitem{draper1998applied}
Norman~R Draper and Harry Smith.
\newblock {\em Applied regression analysis}, volume 326.
\newblock John Wiley \& Sons, 1998.

\bibitem{mitchell1974algorithm}
Toby~J Mitchell.
\newblock An algorithm for the construction of “d-optimal” experimental
  designs.
\newblock {\em Technometrics}, 16(2):203--210, 1974.

\bibitem{sherman1950adjustment}
Jack Sherman and Winifred~J Morrison.
\newblock Adjustment of an inverse matrix corresponding to a change in one
  element of a given matrix.
\newblock {\em The Annals of Mathematical Statistics}, 21(1):124--127, 1950.

\bibitem{sutton2018reinforcement}
Richard~S Sutton and Andrew~G Barto.
\newblock {\em Reinforcement learning: An introduction}.
\newblock MIT press, 2018.

\bibitem{tokic2011value}
Michel Tokic and G{\"u}nther Palm.
\newblock Value-difference based exploration: adaptive control between
  epsilon-greedy and softmax.
\newblock In {\em Annual Conference on Artificial Intelligence}, pages
  335--346. Springer, 2011.

\bibitem{wunder2010classes}
Michael Wunder, Michael~L Littman, and Monica Babes.
\newblock Classes of multiagent q-learning dynamics with epsilon-greedy
  exploration.
\newblock In {\em Proceedings of the 27th International Conference on Machine
  Learning (ICML-10)}, pages 1167--1174. Citeseer, 2010.

\bibitem{ramakrishnan2015many}
Raghunathan Ramakrishnan and O~Anatole von Lilienfeld.
\newblock Many molecular properties from one kernel in chemical space.
\newblock {\em CHIMIA International Journal for Chemistry}, 69(4):182--186,
  2015.

\bibitem{blum2009970}
Lorenz~C Blum and Jean-Louis Reymond.
\newblock 970 million druglike small molecules for virtual screening in the
  chemical universe database gdb-13.
\newblock {\em Journal of the American Chemical Society}, 131(25):8732--8733,
  2009.

\bibitem{rupp2012fast}
Matthias Rupp, Alexandre Tkatchenko, Klaus-Robert M{\"u}ller, and O~Anatole
  Von~Lilienfeld.
\newblock Fast and accurate modeling of molecular atomization energies with
  machine learning.
\newblock {\em Physical review letters}, 108(5):058301, 2012.

\bibitem{buerger2017big}
Philipp Buerger, Jethro Akroyd, Jacob~W Martin, and Markus Kraft.
\newblock A big data framework to validate thermodynamic data for chemical
  species.
\newblock {\em Combustion and Flame}, 176:584--591, 2017.

\bibitem{Linstrom2005nist}
P~Linstrom and W~Mallard.
\newblock Nist chemistry webbook.
\newblock {\em NIST standard reference database}, (69):20899, 2005.

\bibitem{gu2016group}
Geun~Ho Gu and Dionisios~G Vlachos.
\newblock Group additivity for thermochemical property estimation of lignin
  monomers on pt (111).
\newblock {\em The Journal of Physical Chemistry C}, 120(34):19234--19241,
  2016.

\end{thebibliography}


\clearpage
\setcounter{section}{-1}
\renewcommand\thesection{\alph{section}}
\begin{center}
\textbf{\large Supplemental Materials: Designing compact training sets for data-driven molecular property prediction}
\end{center}

\section{S1: Generalized pathway fingerprints for molecular representation}

\begin{figure*}[ht]
\centering
\includegraphics[width=0.6\textwidth]{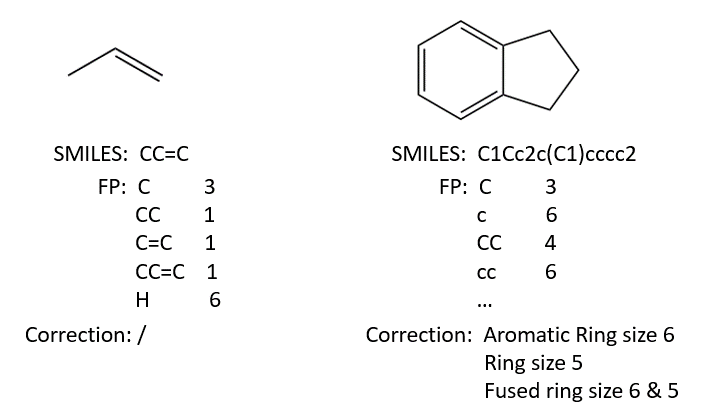}
\caption{Illustrations of linear substructures and correction terms of modified pathway fingerprints on propane and a fused ring molecule. The notation includes (1) SMILES scheme, (2) linear substructures (FP), and (3) correction terms. 
}
\label{FP}
\end{figure*} 

Figure \ref{FP} shows the illustration of the modified pathway fingerprints. The fingerprints consist of two parts, linear substructures (FP) enumerates paths of length one to seven atoms emanating from each atom of the molecule and correction terms (Correction) introduce additional ring information into the fingerprints;  ring information is lumped by correction terms into its presence of aromaticity and its size, while fused ring information is lumped into the size of two adherent rings. The lumped ring information is appended to the feature vector of linear substructures.

\setcounter{section}{-1}

\section{S2: Regularization value selection for the active learning algorithm}

When the initial training set is built or the molecule is updated into the training set, LASSO regression is performed to build the sparse model. The regularization value $\lambda$ determines the number of the selected features and its optimal value is chosen from the range $\left\{\left(2^i\right) | i = i_{upper} : i_{low} : -1 \right\}$, where the upper bond $i_{upper}$ is a predefined large value that allows only a few features to be selected when the training set size is small, and  the lower bond $i_{low}$ is the ten-fold cross validation optimal regularization value based on the initial training set. At each iteration, the regularization value is keeping decreasing from $2^{i_{upper}}$ to $2^{i_{low}}$ by the 2-base logarithmic step and stops at the value where the number of the feature selected is larger than the number of current training molecules or the value reaches the lower bond $2^{i_{low}}$. In the former case, we multiply the value that stops at by 2 and choose the result value for LASSO to allow maximum features to be selected, and in the latter case we choose the lower bound $2^{i_{low}}$.

Sometimes the number of features determined by the regularization value is slightly less than the molecule number in the early iteration steps, which causes severe overfitting and dramatic increase of the RMSE. To prevent such scenario an additional restriction relying on prediction variance function ($x_i^T\left(X^TX\right)^{-1}x_i$) is added that in two consecutive iteration steps, if the selected regularization value leads to a huge increase of the average prediction variance function in the remaining set, we multiply the regularization value by 2 once to select fewer features. In this work, the multiplication is performed if the average prediction variance function is increased more than half of the value in the previous iteration. On the other hand, while more features can be included when the training set size gets large, in several cases the restriction will prevent regularization value getting smaller in later stage even though the average prediction variance function is very small. Thus the restriction is removed to allow more features to be selected after the largest prediction variance function value reaching a small value which indicates a stable model. In this work, the restriction is removed when the largest prediction variance function value is smaller than 1.5.

For some datasets, the lower bound $2^{i_{low}}$ of the regularization range determined by ten-fold cross validation for Max-Min selected initial training set is unreasonably large, which makes feature selected by LASSO too sparse to build an accurate model in the whole iteration. When such a scenario happens, the regularization value decreases to the lower bound within a few iterations, and remains the same for the remaining iterations. One possible explanation is that due to the diversity of the initial training set, during ten-fold cross validation only fragments common to most fragments are selected by LASSO, while other specific fragments contained by a few molecules are likely to be pushed to zero, which leads to a large regularization value being favored. In this work, the scenario happens when applying the algorithm on the surface intermediates set started with Max-Min method for initial training set building, the solution so far is to simply choose the lower bound $2^{i_{low}}$ for Max-Min method selected initial training set from the random selected initial training set instead as it is more uniform with the whole dataset. In principle, we can do another round of cross validation during the middle stage if the too sparse model scenario happens.

\setcounter{section}{-1}

\section{S3: Five-fold Cross validation result for 3 datasets}

\begin{figure*}[ht]
\centering
\includegraphics[width=1\textwidth]{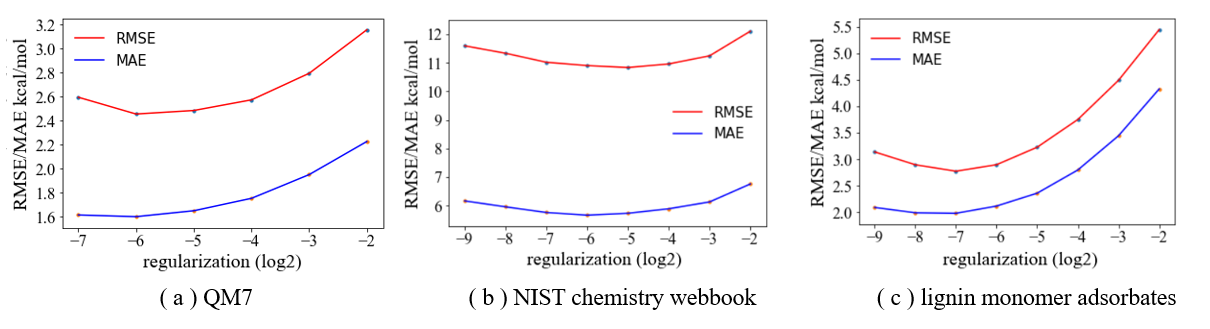}
\caption{Comparison of the average RMSE/MAE with different regularization values for five models during five-fold cross validation for (a) 6.5k gaseous molecules of QM7 dataset with their heats of atomization, (b) 598 hydrocarbons of NIST chemistry webbook with their heats of formation, (c) 591 surface intermediates set with their heat of formation.
}
\label{CV}
\end{figure*} 

Figure \ref{CV} shows the average RMSE/MAE corresponding to different regularization values for five models during five-fold cross validation for three datasets. The lowest RMSE value in each dataset is selected as the optimal value, which is used as the baseline value for the learning rate compare. 

\setcounter{section}{-1}

\section{S4: 10 executions of variance sampling strategy with a randomly selected initial training set for the kernel model}

\begin{figure*}[ht]
\centering
\includegraphics[width=0.6\textwidth]{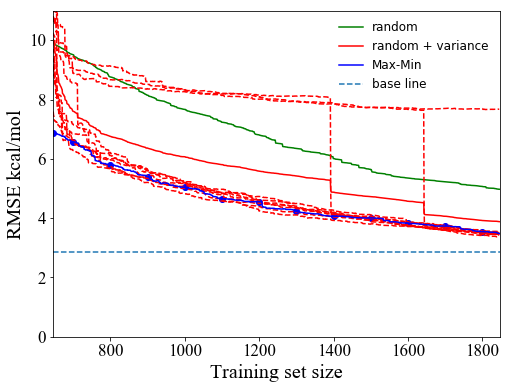}
\caption{Comparison of learning rates for 10 executions of variance sampling strategy started with different initial training sets, each dashed curve represents an execution. The solid curves represent (1) random sampling, (2) pure Max-Min method, and (3) average variance sampling learning rates of 10 executions.
}
\label{flowchart}
\end{figure*} 

Figure \ref{flowchart} shows the learning rates of variance sampling strategy that starts with different randomly selected initial training set determined by different random seeds. 10 executions base on 10 different initial training set are performed. Each dashed curve represents an execution. We can observe that there are three executions that their errors drop perpendicularly after a molecule is updated at a certain step, while for the other executions within a few iterations the curves converge to the kernel distance diversity maximization curve. The result indicates that the randomly selected initial training set may fail to represent some parts of the molecule space due to the absence of a specific type of molecule.

\end{document}